\documentclass[twocolumn,preprintnumbers,amsmath,amssymb]{revtex4}

\usepackage{graphicx}  
\usepackage{epstopdf}
\begin{document}

\title{Effect of doping on the characteristics of infrared photodetectors based on
  van~der~Waals~heterostructures with multiple graphene layers
}
\author{V. Ryzhii$^{1,2,3}$, M. Ryzhii$^4$, 
 V.~Leiman$^5$, V. Mitin$^6$,\\
M. S. Shur$^7$,
 and T.~Otsuji$^1$}
\address{
$^1$ Research Institute of Electrical Communication, Tohoku University,
 Sendai 980-8577, Japan\\
$^2$ Institute of Ultra High Frequency Semiconductor Electronics of RAS,\\
 Moscow 117105, Russia\\
$^3$ Center for Photonics and Infrared Engineering, Bauman Moscow State Technical University, Moscow 111005, Russia\\
$^4$ Department of Computer Science and Engineering, University of Aizu, 
Aizu-Wakamatsu 965-8580, Japan\\
$^5$ Laboratory of 2D Materials' Optoelectronics, Moscow Institute of Physics and Technology, Dolgoprudny 141700, Russia\\
$^6$ Department of Electrical Engineering, University at Buffalo, Buffalo, New York 1460-1920, USA\\
$^7$ Department of Electrical, Computer, and Systems Engineering
and Department of Physics, Applied Physics, and Astronomy, Rensselaer Polytechnic Institute, Troy, New York 12180, USA}

 \begin{abstract} 
\noindent{\bf Keywords:} graphene, van der Waals heterostructure, infrared photodetector.\\ 
We study the operation of infrared photodetectors based on
 van der Waals heterostructures with the multiple graphene layers (GLs) and n-type emitter and collector contacts. The operation of such GL infrared photodetectors (GLIPs) is associated with the photoassisted escape of electrons
from the GLs  into the continuum states in the conduction band of the barrier layers due to the interband photon absorption, the propagation of these electrons and the electrons injected from the emitter across the heterostructure and their collection by the collector
contact.  The space charge of the holes trapped in the GLs provides a relatively strong injection and large photoelectric gain. 
We calculate the GLIP responsivity and dark current detectivity as functions of the energy of incident infrared photons and the structural parameters. It is shown that both the periodic selective doping of the inter-GL barrier layers and the GL doping  lead to a pronounced variation of the GLIP spectral characteristics, particularly 
near the interband absorption threshold, while the doping of GLs solely results in a substantial increase in the GLIP detectivity. The doping "engineering" opens wide opportunities for the optimization of GLIPs for operation in different parts of  radiation spectrum from near infrared to terahertz.  
\end{abstract} 
 
\maketitle

\newpage

\section{Introduction}
The gapless energy spectrum of graphene layers (GLs)  enables their use in the interband  detectors of  infrared   radiation~(see, for example,~\cite{1,2,3,4,5}).
The incorporation of  the GLs  into 
 the van der Waals (vdW) heterostructures based on such materials as 
hBN, WS$_2$, InSe, GaSe,  and similar materials~\cite{6,7,8,9,10,11,12,13,14,15}    can enable the creation
of  novel GL infrared  photodetectors (GLIPs) with the improved characteristics.

Recently, we proposed and evaluated the IR detectors 
using the vdW) heterostructures with the GLs clad by the widegap barrier layers - GL infrared photodetectors (GLIPs)~\cite{16,17}. The GLs serve as photosensitive elements, in which electron-hole pairs are generated due to the interband absorption of IR radiation. The photogenerated electrons tunnel from the GLs through the barrier top to the continuum states in the barrier layers and support the terminal current. The photogenerated holes, which  are confined in the GLs,  form the space charge. The space charge is determined by the balance of the photogeneration and capture of the electrons propagating above the barriers. Figures~1 and 2  show the GLIP schematic view and  the fragment of the device band diagram with the indicated main electron processes (the photoexcitation and the tunneling from and capture into the GLs).
The space charge affects  the electric ﬁeld at the device emitter and, therefore, controls the injected electron current.
In the devices based on the heterostructures with  a low efficiency of the electron capture into the GLs, the injected current can markedly exceed the current created by the photoexcited electrons. This  provides a relatively high photoelectric gain and detector responsivity. 
The rates of the escape of the photoexcited and thermalized electrons from the GLs and the capture of the electrons propagating across the barrier layers strongly depend of  the potential profile near the GLs.
The doping of the barrier layers, in particularly, the selective doping using the delta layers of donors and acceptors as shown in Figs.~1 and 2 (which is called the "dipole" doping~\cite{18}) can markedly modify this profile resulting in the appearance of the "tooth" adjacent to each inner GL at the donor sheet side. 
The barrier doping was effectively used also  in the unitravelling-carrier (UTC) photodiodes to reinforce  the injection
of the electrons photogenerated in the emitter of these devices~\cite{19}.
 The doping of the GLs  can also lead shift of the Fermi level in the GLs with respect to the Dirac level. The latter affects the spectrum of the electron photoexcitation, the escape rate of the thermalized electrons, and the capture processes.

 The characteristics of the GLIPs considered previously~\cite{16,17} are primarily predetermined by the electron affinities of GLs and barrier materials.
In this paper, we show that the proper doping of the barrier layers  and GLs by acceptors and donors can pronouncedly modify the GLIP characteristics and result in  increase in the  GLIP responsivity and detectivity, particularly, in the  low-energy part of the infrared spectrum.

\section{Structure of GLIPs and their operation principle}

\begin{figure}[t]
\centering
\includegraphics[width=6.0cm]{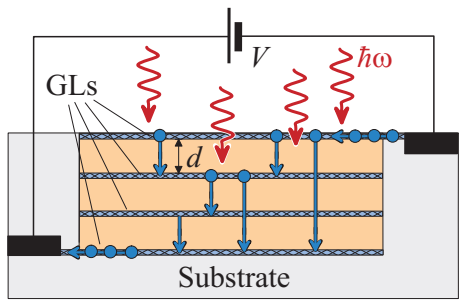}
\caption{Schematic view of the GLIP heterostructure. Horizontal arrows correspond to electron flow from the emitter 
along the emitter GL and along the collector GL to the collector contact. Vertical arrows indicate
 flow of the electrons injected from the emitter GL and photoexcited from the inner GLs across one, two, or more inter-GL barriers before being captured.}
\end{figure}

\begin{figure}[t]
\centering
\includegraphics[width=7.0cm]{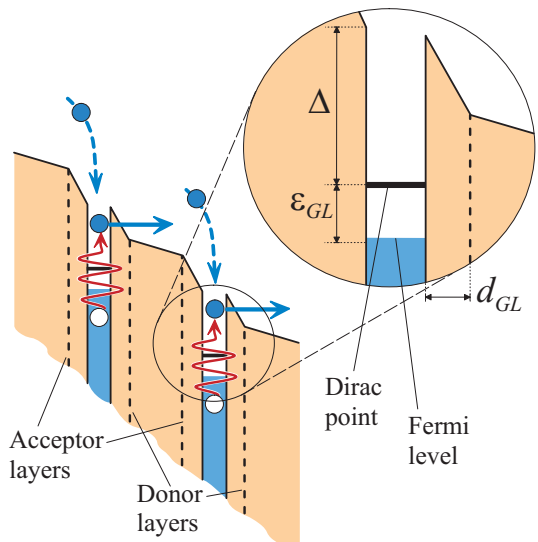}
\caption{A fragment of the GLIP band diagram of a GLIP with  barrier layers doped by acceptors and donors ("dipole" doped) and GLs doped by acceptors. Wavy, solid, and dashed  arrows  indicate  the processes of the electron photoexcitation, tunneling, and capture, respectively.  The inset shows the barrier "tooth" and its parameters.}
\end{figure}

The GLIP under consideration consists of the GL-vdW heterostructure, which comprises  $N = 1,2,3,...$ GLs  clad by
 the barrier layers and  the two the emitter and collector GLs (the top and bottom GLs, respectively). The latter GLs are doped by donors to provide their sufficiently high lateral conductivity. In contrast with our previously considered GLIPs,~\cite{16,17} we assume that the inter-GL barriers are selectively doped by acceptors and donors (as shown in Fig.~2) with equal densities $\Sigma_B$.
The inner GLs can also be doped by acceptors with the density $\Sigma_{GL}$.

To provide the localization of the photoexcited holes, the valence band offset, $\Delta_V$, 
between  the GLs and  barrier layers is larger than the conduction band offset $\Delta$ (i.e., $\Delta < \Delta_V$ or $\Delta \ll \Delta_V$). A sufficiently strong  dc bias voltage $V$ is applied between the contact GLs.

The GLIP operation  is associated with the following processes~\cite{16,17}: 
(1) the photoexcitation of the electron-hole pairs in the GLs due to the interband radiative transitions; 
(2)  the tunneling injection of the thermalized electrons from the ground states in the GLs and 
the escape of the photoexcited electrons from their  excited states followed by  the propagation across the barrier layers;
(3) the electron capture from the continuum states above the inter-GL barriers into the inner GLs. 

\section{Equations of the model}

Generalizing the results of the recent calculations~\cite{16,17}, the density of the current across the GLIP caused by the incident infrared radiation with the intensity $I$ (inside the device) and photon energy $\hbar\omega$
can be presented as

\begin{eqnarray}
j_{photo} = \frac{e\beta_{\omega}\theta_{\omega}\,I}{(\gamma_E^{3/2} + N)}
\biggl[\gamma_E^{3/2} +   \displaystyle\frac{(1 - \beta_{\omega})[1 - (1 -\beta_{\omega})^{N}]}{p\beta_{\omega}}\biggr]
\end{eqnarray}\label{eq1}
with
\begin{equation}\label{eq2}
\beta_{\omega} = \frac{\beta\exp\biggl(\displaystyle\frac{\hbar\omega}{2T}\biggr)}{2\biggl[\cosh\biggl(\displaystyle\frac{\varepsilon_{GL}}{T}\biggr) + \cosh\biggl(\displaystyle\frac{\hbar\omega}{2T}\biggr)\biggr]}
\end{equation}
and
\begin{equation}\label{eq3}
\theta_{\omega} = \frac{1}{1  +  \displaystyle\frac{\tau_{esc}}{\tau_{relax}}\exp\biggl( \frac{\eta_{\omega}^{3/2}E_{tunn}}{E_{GL}}\biggr)}.
\end{equation}
Here the  factor depending on $\hbar\omega$  in Eq.~(2) reflects the Pauli exclusion principle,
the quantity
 $\theta_{\omega}$  is
 the probability of the escape from the GL of the electrons photoexcited owing to the interband absorption of the photons with the energy $\hbar\omega$,
$\beta = \pi\alpha/\sqrt{\kappa_{B}}$, $\alpha = e^2/\hbar\,c$ is the fine structure constant, $e$ and $\hbar$ are
the electron charge and the Planck constant, $\sqrt{\kappa_B}$ is the barrier material refractive index,
$T$ is the temperature, $\varepsilon_{GL}$ is  Fermi energy in the inner GLs counted from the Dirac point, $p$ is the capture efficiency~\cite{20,21,22,23,24,25} (which in the heterostructures under consideration can be very  small: $p \ll 1$~\cite{26}), $N$ is the number of the inner GLs, and
$\gamma_E  = (\Delta - \varepsilon_E)/\Delta< 1$, where $\varepsilon_E$ is the electron Fermi energy in the emitter GL.   The parameter $\gamma_E $ plays the role   of the emitter ideality parameter.
It depends  on the features of the electron injection from the emitter into the GLIP heterostructure bulk~\cite{22,23,24,25,26,27}.  For the "ideal" emitter contact $\gamma_E = 0$, and the electric field in the near-emitter barrier is close to zero. This corresponds to the situation when the emitter provides the injection of such an amount of electrons which is dictated by the conditions in the device bulk.

The probability, $\theta_{\omega}$, of the photoecited electrons escape from the GLs is determined
by the ratio of the try-to- escape time $\tau_{esc}$ and the electron energy relaxation time $\tau_{relax}$, and by the tunneling exponent. The latter depends on the energy (with respect to the barrier top)  of the photoexcited electrons via the factor $\eta_{\omega} = (\Delta - \hbar\omega/2)/\Delta$ if $(\Delta - \Delta_{GL})  \leq\hbar\omega \leq 2\Delta$ and $\eta_{\omega} = 0$ if  $\hbar\omega > 2\Delta$,  the characteristic "tunneling" field  
$E_{tunn} = 4\sqrt{2m} \Delta^{3/2}/3e\hbar$~\cite{28},
(where $m$ is the electron effective mass in the barrier layer), the height of the barrier "tooth" adjacent to the GL $\Delta_{GL}$, which is determined by  the real electric field in the barriers at the inner GLs $E_{GL}$, Considering the doping of the barrier layers,  we obtain the following formulas for the electric fields near the GLs $E_{GL}$ and in the bulk of the barriers $E_B$: 

\begin{equation}\label{eq4}
E_{GL} = \frac{V}{(\gamma_E^{3/2} + N)d} + \frac{V_B}{d}\biggl(1 - \frac{2d_{GL}}{d}\biggr),
\end{equation}
\begin{equation}\label{eq5}
E_{B} = \frac{V}{(\gamma_E^{3/2} + N)d} - \frac{V_B}{d}\biggl(\frac{2d_{GL}}{d}\biggr),
\end{equation}
with 
$V_B = 4\pi\,e\Sigma_Bd/\kappa_B$ and $\Delta_{GL} = E_{GL}d_{GL}$,
where $d$ is the barrier layer thickness and $d_{GL} $ is the spacing between the GLs and the donor sheets
see the inset in Fig.~2).
The barrier doping effectively increases the rate of the photoexcited electrons tunneling rate when 
the "tooth" height $\Delta_{GL} $ is sufficiently large, in particular, if $\Delta_{GL} \simeq \Delta$. For the definiteness, the latter relation is assumed in the following.
The Fermi energy of holes in the inner GLs at not to high temperatures can be expressed via the acceptor and hole density $\Sigma_{GL}$ as
$\varepsilon_{GL} = \hbar\,v_W\sqrt{\pi\Sigma_{GL}}$,where $v_W \simeq 10^8$~cm/s, 

\begin{figure}[t]
\centering
\includegraphics[width=7.0cm]{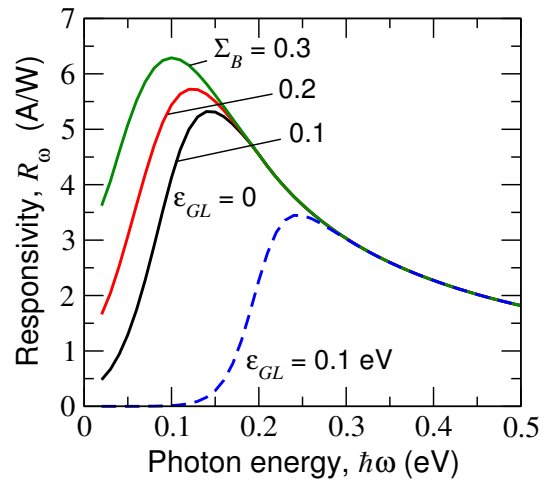}
\caption{
Spectral characteristics of GLIPs with $\Delta = 0.1$~meV,  five undoped  GLs ($\Sigma_{GL} = 0$), and  different donor and acceptor densities in the barriers $\Sigma_B$ (in units $10^{12}$~cm$^{-2}$ ) at $T = 100$~K. Dashed line corresponds to
 GLIP with doped GLs ($\varepsilon_{GL} = 0.1$~eV, $\Sigma_{GL} = 0.8\times10^{12}$~cm$^{-1}$) - curves for different barrier doping are undistinguished.
 }
\end{figure}

\begin{figure}[t]
\centering
\includegraphics[width=7.0cm]{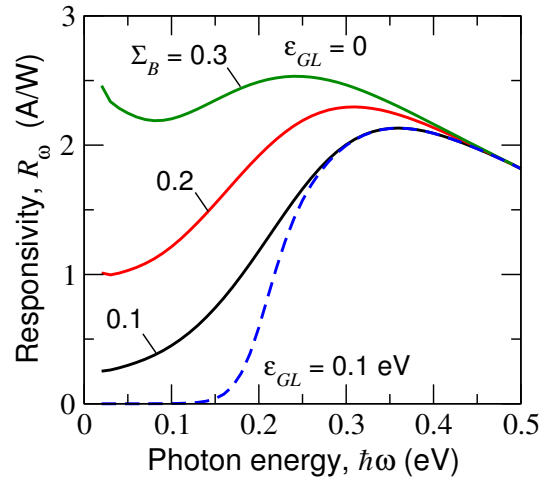}
\caption{The same as in Fig.~3 but for GLIPs with $\Delta = 0.25$~eV.}
\end{figure}

\section{GLIP responsivity}

Equation~(1) for the GLIPs with  not too large (realistic) number of the GLs in the GLIP part ($N \ll \beta_{\omega}^{-1}$), considered in the following,  can be somewhat simplified leading to the following formula for the GLIP responsivity $R_{\omega}  \propto j_{photo}/\hbar\omega\,I$:

\begin{eqnarray}\label{eq6}
R_{\omega}\simeq  R
\frac{\displaystyle\biggl(\frac{\Delta}{2\hbar\omega}\biggr)\frac{N}{(\gamma_E^{3/2} + N)}}{\biggl[1  +  \displaystyle\frac{\tau_{esc}}{\tau_{relax}}\exp\biggl( \frac{\eta_{\omega}^{3/2}E_{tunn}}{E_{GL}}\biggr)\biggr]}\nonumber\\
\times\frac{\exp\biggl(\displaystyle\frac{\hbar\omega}{2T}\biggr)}{\biggl[\cosh\biggl(\displaystyle\frac{\varepsilon_{GL}}{T}\biggr) + \cosh\biggl(\displaystyle\frac{\hbar\omega}{2T}\biggr)\biggr]}
\end{eqnarray}
with 
\begin{equation}\label{eq7}
R = \frac{e\xi\beta}{p\Delta}.
\end{equation}
Here the factor $\xi$ is determined by  conditions of reflection of the incident radiation from the GLIP top interface~\cite{14}.
Equation~(6) turns to  that derived and used previously for GLIPs with the undoped barriers and inner GLs~\cite{16,17}  at $\varepsilon_{GL} = 0$ and $\Sigma_B = 0$ when $E_{GL}$ replaced by $V/(\gamma_E^{3/2} + N)d$.
For example, setting $\xi = 1$, $p = 0.01$, $\kappa_B = 5$, and $\Delta = 0.1 - 0.5$~eV, for the characteristic responsivity  we obtain $R \simeq (2 - 10) $~A/W.

The value of the GLIP responsivity given by Eq.~(5) with Eq.~(6) corresponds to the photoconductive gain
$g = [p(\gamma_{E}^{3/2} + N)]^{-1} \simeq (pN)^{-1}$~(compare with Refs.~\cite{20,21,25,29}). The origin of this gain is associated with the accumulation of the charges formed in the GLs by the photogenerated holes. The latter is due to a much more effective  confinement of the photogenerated holes than the photogenerated (photoexcited) electrons that stems from
the condition $\Delta < \Delta_V$ accepted in Sec.~II. If this condition is violated, the escape probability of the photogenerated holes from the GLs can become rather high leading to vanishing of the phoconducting gain effect.

Figures~3 and 4 show the  responsivity calculated using Eq.~(6) with Eqs.~(4) and (5) as a function of the photon energy  for the GLIPs with different barrier heights $\Delta$  and different doping levels of the inter-GL barriers and GLs.
 For the definiteness, the following general parameters are assumed: $N = 5$, $\xi = 1$,  $p = 0.01$, $\gamma_E  \ll 1$, $d = 10$~nm, $d_{GL} = 2$~nm, $\tau_{esc}/\tau_{relax} = 0.1$, $T = 100$~K, and $U =V/dE_{tunn} = 0.5$.
At $m = (0.14 - 0.28)m_0$ ($m_0$ is the mass of bare electron) and $d = 10$~nm. The value $U = 0.5$ corresponds to
$V/N = 0.026 - 0.038$~V at $\Delta = 0.1$~eV and $V/N = 0.11 - 0.15$~V at $\Delta = 0.25$~eV.

One can see that an increase in the barrier doping level, which results in higher tunneling transparency of the barrier for the photoexcited electrons and, hence, higher probability of their escape, leads to a substantial increase in the responsivity at relatively low photon energies ($\hbar \omega < 2\Delta$). The responsivity of the GLIPs with smaller $\Delta$ is higher than that of the GLIPs with larger $\Delta$ in the low photon energy range (compare the curves in Figs.~3 and 4). An increase in the responsivity at relatively low photon energies exhibited by the curve for $\Sigma_B = 0.3\times 10^{12}$~cm$^{-2}$ in Fig.~4 is   attributed to the factor $1/\hbar\omega$ in Eq.~(6) (see also a comment in Sec. ~VI). 
Marked values of the responsivity in the range $\hbar\omega \lesssim 0.05$~eV (about several A/W, as seen from Figs.~3 and 4) imply that the GLIPs with properly doped barrier layers can operate not only in near- and 
mid-infrared spectral ranges but also in the terahertz range.

The GL doping by acceptors also modifies the responsivity 
spectral dependence: its increase  (and, therefore, increase in the Fermi energy $\varepsilon_{GL}$)gives rise to a marked shift of this dependence toward higher photon energies (compare the solid and dashed lines in Figs.~3 and 4).
In the case of doped GLs, the barrier doping weakly affects the spectral dependence in question (the curves corresponding to different values of $\Sigma_B$ and $\varepsilon_{GL} = 0.1$~eV are practically merged.

In principle, the temperature smearing of the electron energy distributions in the GLs somewhat affects
the photon absorption probability at $\hbar\omega \simeq 2(\Delta + \varepsilon_{GL})$
due to the degeneracy of the electrons system near the Fermi level. However,
the  variation of the temperature (in the range $T = 50 - 200$ only slightly changes the above spectral dependences.

\begin{figure*}[t]
\centering
\includegraphics[width=15.0cm]{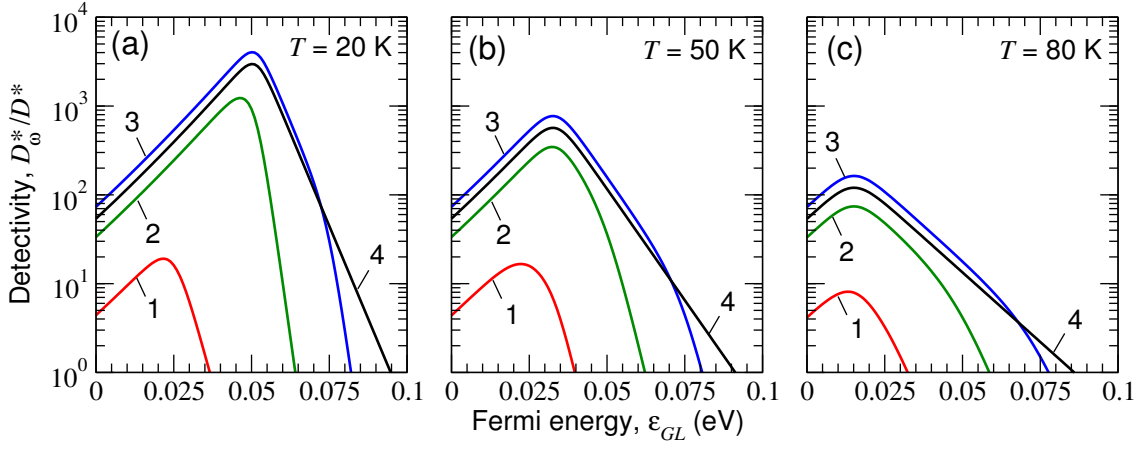}
\caption{GLIP detectivity  for $\Delta = 0.1$~eV and undoped barriers ($\Sigma_B = 0$) as a function of the 
 Fermi energy $\varepsilon_{GL}$ (acceptor density in GLs $\Sigma_{GL}$) for different photon energies 
 $\hbar\omega$ at (a) $T = 20$~K, (b) $T = 50$~K, and (c) $T = 80$~K: 1 - $\hbar\omega = 0.05$~eV, 2 - $0.1$~eV, 3 - $\hbar\omega = 0.15$~eV, and 4 - $0.25$~eV.  
 }
\end{figure*}

\begin{figure*}[t]
\centering
\includegraphics[width=11.0cm]{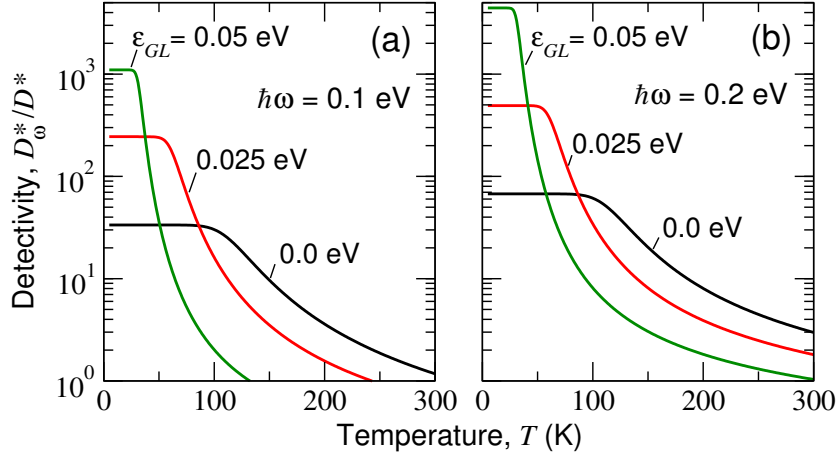}
\caption{Temperature dependences of detectivity of GLIPs with $\Delta = 0.1$~eV, undoped barriers and different
 Fermi energies $\varepsilon_{GL}$ for (a)  $\hbar\omega = 0.1$~eV and (b)  $\hbar\omega = 0.2$~eV.}
\end{figure*}

\begin{figure*}[t]
\centering
\includegraphics[width=11.0cm]{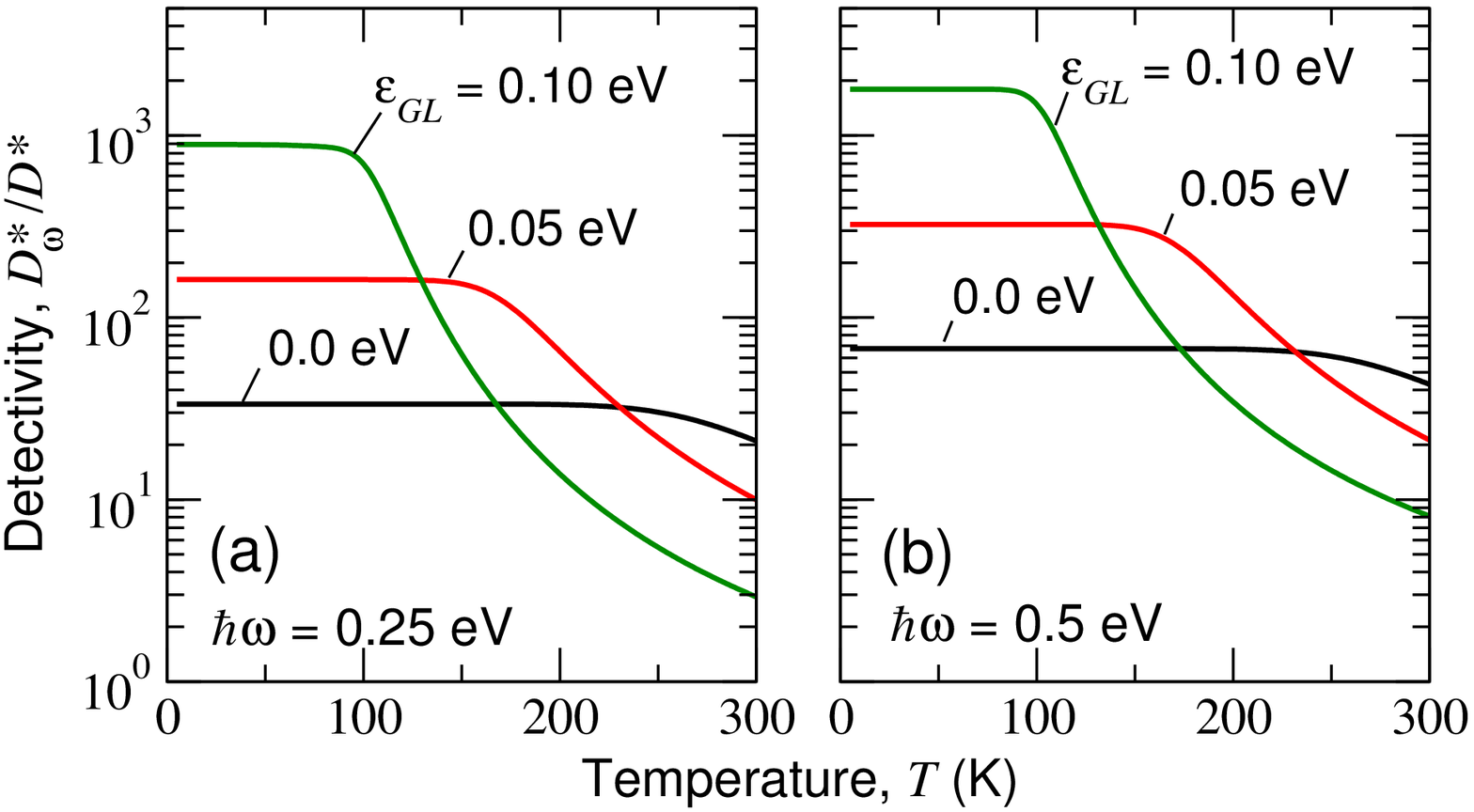}
\caption{ The same as in Fig.~6 but for  $\Delta = 0.25$~eV and (a)  $\hbar\omega = 0.25$~eV and (b) $\hbar\omega = 0.5$~eV.}
\end{figure*}

\section{GLIP detectivity}

The dark current limited detectivity is usually determined as

\begin{equation}\label{eq8}
D^{*}_{\omega} = \frac{R_{\omega}}{\sqrt{4egj_{dark}}},
\end{equation}
where  $g$ is the photoconductive gain which was introduced in Sec.~IV.
The dark current in the GLIPs is determined by the tunneling of the thermalized electrons from the GLs (amplified by the electron injection from the emitter GL).  One can assume that  the main contribution to this tunneling
is provided by the electrons with the energies closed to the Fermi level. Due to the specific features of the tunneling barrier shape, the tunneling exponent depends of the barrier parameters $\Delta_B$ and $d$.
Considering this and generalizing the pertinent equations obtained for the GLIPs with the undoped barriers and  inner GLs,
 one can use the following relation for the dark current:

\begin{equation}\label{eq9}
j_{dark} = \frac{j_{max}f_E}{p}\exp \biggl[- \frac{(\eta^{3/2} + F)E_{tunn}}{E_{GL}}\biggr],
\end{equation}
where for $\Delta_{GL} = \Delta$

\begin{equation}\label{eq10}
\eta = 1 + \frac{\varepsilon_{GL}}{\Delta}  = 1 + \frac{\hbar\,v_W\sqrt{\pi\Sigma_{GL}}}{\Delta},
\end{equation}

\begin{eqnarray}\label{eq11}
F=  \biggl(\frac{E_{GL}}{E_B} - 1\biggr)\biggl(\frac{\varepsilon_{GL}}{\Delta}\biggr)^{3/2}\nonumber\\
 =  \frac{(d/d_{GL})}{\displaystyle\frac{V}{(\gamma^{3/2} + N)V_B} -1}\biggl(\frac{\hbar\,v_W\sqrt{\pi\Sigma_{GL}}}{\Delta}\biggr)^{3/2},
\end{eqnarray}
$j_{max}$ is the maximum current density which can be extracted from the emitter GL, and $f_E$ is the pre-exponential factor, which depends on the emitter ideality factor $\gamma_E$. Disregarding for brevity the effects associated with the emitter nonideality (analyzed previously,~\cite{16,17}), we put in the following $\gamma_E = 0$ and $f_E =1$. 

At elevated temperatures, the thermionic escape of electrons from the GLs can also contribute to the dark current.
The pertinent dark current density can be presented as

\begin{equation}\label{eq12}
j_{dark}^{(therm)} = \frac{cj_{max}}{p}\exp\biggl( - \frac{\Delta_{therm}}{T}\biggr),
\end{equation}
where the quantity $\Delta_{therm}$ plays the role of the thermionic activation energy for the thermalized electrons in the GLsand $c \sim 1$.
Generally, $\Delta - \Delta_{GL}+ \varepsilon_{GL} \lesssim \Delta_{therm} \lesssim \Delta + \varepsilon_{GL}$.

Considering for simplicity the interpolation formula for the net dark current density in which the contributions given by Eqs.~(9) and (12) are summarized, we arrive at 
the following equation for the dark current limited GLIP detectivity:

\begin{eqnarray}\label{eq13}
D^{*}_{\omega} = D^{*} \frac{\biggl(\displaystyle\frac{\Delta}{2\hbar\omega}\biggr)}{(1 + \Theta_{therm})^{1/2}}\frac{\exp \biggl[\displaystyle\frac{(\eta^{3/2} + F)E_{tunn}}{2E_{GL}}\biggr]}{\biggl[1  +  \displaystyle\frac{\tau_{esc}}{\tau_{relax}}\exp\biggl( \frac{\eta_{\omega}^{3/2}E_{tunn}}{E_{GL}}\biggr)\biggr]}\nonumber\\
\times\frac{\exp\biggl(\displaystyle\frac{\hbar\omega}{2T}\biggr)}{\biggl[\cosh\biggl(\displaystyle\frac{\varepsilon_{GL}}{T}\biggr) + \cosh\biggl(\displaystyle\frac{\hbar\omega}{2T}\biggr)\biggr]}
\end{eqnarray}
with
\begin{equation}\label{eq14}
 D^{*} = \frac{\sqrt{N}e\xi\beta}{\sqrt{4ej_{max}}\Delta}.
\end{equation}
Here 
\begin{equation}\label{eq15}
\Theta_{therm} = c\exp\biggl[\frac{(\eta^{3/2} + F)dNE_{tunn}}{E_{GL}} - \frac{\Delta_{therm}}{T}\biggr]
\end{equation}
is the quantity characterizing the relative contribution of the tunneling and thermionic processes. 
For the GLIPs with undoped barriers setting $E_{GL} = V/dN$, $F = 0$, and $\Delta_{therm} \simeq \Delta + \varepsilon_{GL}$, Eq.~(15) yields 

\begin{equation}\label{eq16}
\Theta_{therm} \simeq \exp\biggl(\frac{\eta^{3/2}NdE_{tunn}}{V} - \frac{\Delta + \varepsilon_{GL}}{T}\biggr).
\end{equation}

Assuming $N = 5$ and $j_{max} = 1.6\times (10^5 - 10^6)$~A/cm$^2$, at the same other parameters as in the above estimate of the characteristic responsivity $R$, we obtain $D \simeq (0.5 - 7)\times 10^5$~cm$\sqrt{\rm{Hz}}$/W.
Due to a large first exponential factor in Eq.~(13), the real detectivity $D_{\omega}^{*} \gg D^{*}$. The GLIPs with a larger number of the inner GLs $N$ can exhibit higher values of the  dark current limited detectivity (because $D^{*} \propto \sqrt{N}$~\cite{29}).
The values of $j_{max}$ used here correspond, in particular, to the electron density in the emitter GL $\Sigma_E = 10^{12}$~cm$^{-2}$ and the try-to-escape time $\tau_{esc} = 10^{-13} - 10^{-12}$~s.

According to Eq.~(13), the spectral dependence of the detectivity repeats that of the responsivity (shown, in particular, in. Figs.~3 and 4). The dipole doping of the barrier layers leads to an increase in the GLIP responsivity
(primarily in the range of relatively low photon energies) but simultaneously  to an increase of the dark current and, hence, a drop of the detectivity. The doping of GLs by acceptors, which  modifies the spectral characteristics, promotes the dark current lowering and a rise of the detectivity. In principle, carefully choosing the levels of both type of doping, one can expect the optimal relation between  the responsivity and the detectivity. However, taking into account 
that realization of both types of doping in one devices can markedly complicate its fabrication,   we restrict ourselves by considering the detectivity of the GLIPs  with the doping of the GLs only.
Therefore, we focus on the GLIP detectivity as a function of the GL doping and the temperature assuming that the barrier layers are undoped.
 
Figure~5  shows the detectivity of the GLIPs with undoped barrier layers calculated using Eqs.~(13) and (16)  as a function the Fermi energy $\varepsilon_{GL}$ (which is determined by the acceptor density in the GLs) for different temperatures.
Figures~6 and 7 demonstrate how  the detectivity of the GLIPs with different barrier height $\Delta$ and  Fermi energy $\varepsilon_{GL}$ (i. e., different acceptor densities in the GLs $\Sigma_{GL}$) at different photon energy $\hbar\omega$ varies with increasing temperature $T$. 
 We set  $N = 5$, $\tau_{esc}/\tau_{relax} = 0.1$, and  $U = 0.5$. The Fermi energy $\varepsilon_{GL}$
 changes from zero to 0.1~eV  in the acceptor density range $\Sigma_{GL} = (0 - 8)\times 10^{12}$~cm$^{-2}$.

As seen from Fig.~5, the detectivity $D_{\omega}^*$ is a nonmonotonic function of the Fermi energy $\varepsilon_{GL}$ and the acceptor density $\Sigma_{GL}$ in the GLs with  a pronounced maxima at 
certain values $\varepsilon_{GL}$ and  $\Sigma_{GL}$. A pronounced increase in the detectivity is attributed to an increase in the barrier height  for the thermalized electrons $\Delta + \varepsilon_{GL}$
with rising doping level (see Fig.~2) leading to a diminishing of the tunneling and thermionic electron escape and, consequently, to a weaker dark current. 

A steep detectivity roll-off at increased acceptor densities is associated with the Pauli principle leading to an abrupt drop of the photon absorption and, hence, the responsivity  when $\varepsilon_{GL}$ becomes close or larger than $\hbar\omega/2 $. Some difference in the steepness of the detectivity roll-off seen in Figs.~5(a) - 5(c) is due to a stronger  smearing of the Fermi-Dirac distribution in the GLs  at higher  temperatures.
The dependences shown in Figs.~5 - 7 indicate a marked decrease in the detectivity maximum with rising temperature. This is explained by an increase in the   role  of  thermionic processes at elevated temperatures.

One needs to point out that the values of the detectivity $D_{\omega}^{*}$ at certain values of the  Fermi energies
(acceptor densities) can be rather high. Taking into account the values of $D^{*}$ obtained in the above estimate,
for the $D_{\omega}^{*}$ maximum we find max$D_{\omega}^{*} > 10^9$~cm$\sqrt{\rm{Hz}}$/W.

One can see from Figs.~6 and 7 that the detectivity being a flat  function of the temperature steeply drops at $T$  exceeding a certain temperature: 

\begin{equation}\label{eq17}
T_{therm} \sim \frac{\Delta^{3/2}}{(\Delta + \varepsilon_{GL})^{1/2}}\biggl(\frac{V}{NdE_{tunn}}\biggr).
\end{equation}
 This is associated with the inclusion of the thermionic contribution to the electron escape from the GLs. Although the enhancement of the GL doping results in a pronounced rise in the detectivity, it leads to a shrinking of the temperature range where $D_{\omega}^{*}$ [and crossings of the curves in Figs.~(6) and (7)].
Indeed, Eq.~(17) yields $T_{therm} \propto \Delta^{3/2}/(\Delta + \varepsilon_{GL})^{1/2}$, i.e., a decreasing
$T_{therm}$ versus $\varepsilon_{GL}$ relation.

\section{Discussion}

As follows from the above results, the GLIPs  under consideration can have high responsivity and detectivity in
near and far IR spectral  ranges. The GLIPs with $\Delta \sim 0.1$~eV can operate also in the terahertz range
$\hbar\omega \sim 0.025 - 0.05$~eV ($\hbar\omega/2\pi \sim 6 - 12$~THz), exhibiting  reasonable values of the responsivity and detectivity.

In Sec.~IV, 
we have omitted the analysis of the responsivity  spectral behavior in the range of very low photon energies, because the interband absorption in this range can be complicated by
the smearing of the carrier spectrum spectrum, fluctuations of potential profile (existence of the electron-hole puddles), and formation of a narrow energy gap (due to specific doping or substrate properties) ~\cite{30,31,32,33,34,35} - factors which are not described by Eqs.~(2) and (3) and, hence, are beyond our device model.

The  barrier height for the thermalized electrons in the GLIPs with doped GLs  is equal to $\Delta_{therm} = \Delta^{(doped)} + \varepsilon_{GL}$. This height determines both the tunneling and thermionic dark currents.
Comparing the dark currents in a GLIP with the acceptor doped GLs and with that in a GLIP with undoped GLs but with a higher barrier ($\Delta^{(undoped)} = \Delta_{therm} > \Delta^{(doped)}$),
one can find that  these currents are  equal to each other or, at least, of the same order of magnitude.
In contrast, taking into account that
the barrier heights for the electrons photoexcited in the doped and undoped GLs are equal to $ \Delta^{(doped)} - \hbar\omega/2$ and $ \Delta^{(undoped)} - \hbar\omega/2$, respectively, we conclude that the escape rate of the photoexcited  electrons and, hence, the GLIP responsivity   in the former case is larger than that in the latter case (because $\Delta^{(doped)} < \Delta^{(undoped)}$). Thus the GL doping by acceptors offers  better GLIP performance
in comparison with the GLIPs without doping of the  GLs but elevated barrier height (larger difference in the GL and barrier material affinities).

Calculating both the photocurrent and the dark current in the GLIPs with the doped barriers we have disregarded the effect of the donor and acceptor spatial fluctuation in the device plane. These fluctuations can lead to pronounced fluctuation of the electric field $E_{GL}$ and, consequently, the tunneling current created by the photoexcited and thermalized electrons (see, for example, Ref.~\cite{36,37}). As for the photocurrent and, hence, the GLIP responsivity
the fluctuations in question promote an increase in these quantities. This implies that the values of the GLIP responsivity can somewhat exceed those obtained above. Since, considering, the detectivity, focused on  the GLIPs with undoped barrier layers, the problem of doping fluctuation is out of the scope of this work.

Since the photoexcited and injected electrons acquire kinetic energy propagating across the device under the electric field, they can be hot. If the electron energy relaxation length $L_{\varepsilon} = v_d\tau_{\varepsilon}$, where $v_d$ and  $\tau_{\varepsilon}$ are the electron drift velocity and the energy relaxation time, respectively, exceeds the heterostructure period $d$, the electron effective temperature $T_{eff}$ is mainly determined by the applied bias voltage $V$~\cite{38}. 
An increase in $V$ leads to a rise of $T_{eff}$ and to a drop of the capture efficiency $p$~\cite{26} (see also references therein).
Since the responsivity $R_{\omega} \propto 1/p$ [see Eqs.~(6) and (7)], the electron heating promotes higher values of the  responsivity.
As demonstrated by the particle Monte Carlo  modeling of the electron capture into quantum wells (QWs) in heterostructures (albeit  made of the standard material) with doped barriers~\cite{39}, the doping affecting the potential profile
in the barrier layers can result in a somewhat steeper drop of the capture efficiency with increasing voltage.
 The hot propagating electrons can provide a heating of the carriers localized  in the GLs
enhancing the electron thermionic escape. Apart from this, some fraction of the energy of the absorbed photons goes to
the carrier heating~\cite{17}.  
This can lead to a decrease in the GLIP detectivity. However, one might expect that at a small  capture efficiency and not too high radiation intensities, the negative impact of the heating is not to strong. 
The electric field across the GLs, modifying the  wave functions of the photoexcited and thermalized electrons and, hence, the try-to-escape time, can lead to an increase in both the photocurrent and the dark current.
This promotes higher values of the GLIP responsivity, but can add complexity to the voltage dependences of the GLIP detectivity. The consideration of the latter,  as well as  more rigorous treatment of the thermo-assisted tunneling, require a generalization of the GLIP model that  is beyond the scope of this work.

Generally, the selection of materials for the GLIPs from a wide variety of them is a matter of using of a proper band alignment (see, for example,~Refs.~\cite{14,40}).  Several already fabricated and experimentally studied devices using  the vdW heterostructures with the GLs, which can serve as the reference points for the GLIP realization, were reported recently~\cite{41,42,43,44,45,46,47}.
The value $\Delta = 0.1$~eV, which was used in some of the above calculations, might, in particular,  correspond to the GLIPs with the InSe 
barrier layers~\cite{8,9,48}.
The GLIPs with relatively low barriers able operate in the THz range can be based on the Oxide family materials 
with the electron affinity close to that in GLs (for example, RuO$_2$ and TiO$_2$~\cite{49}).

The comparison of the GLIP characteristics with the characteristics of photodetectors, using similar operation principles, namely with intersubband quantum well infrared photodetectors (QWIPs)~\cite{25,29} shows the following advantages of the former:\\
(i) A higher responsivity due to both  higher probability of the electron photoexcitation associated with the use of the interband transitions in the  GLs and the  intraband (intersubband) transitions in the QWs;\\
(ii) A higher detectivity associated with  higher responsivity and weaker dark current (due to a larger activation energy);
(iii) Sensitivity to normally incident radiation (because of the use of the interband transition), so there is no need in special coupling structures;\\
(iv) No need in the  GL doping, although such a doping, as shown above, provides an opportunity to vary the GLIP characteristics, in particular, enhancing the GLIP performance;\\ 
(v) Possibility of the GLIP operation in the range $\hbar\omega \simeq 0.025 - 0.05$~eV  ($\omega/2\pi \simeq 6 -12$~THz), where using more conventional materials (e.g. III-V compounds) is hindered by optical phonon absorption.

\section{Conclusions}
We studied the effect of the barrier layer and GL doping on the responsivity and detectivity of the GLIPs 
intended for the detection of infrared radiation. Using the developed device model, we demonstrated that
the doping can result in a substantial modification of the spectral characteristics and enhancement of the GLIP responsivity and detectivity. The obtained results can be used for the characteristic  optimization of  
the GLIP operating in different parts of the  spectrum, including the terahertz range.

\section{Acknowledgments}

 The authors are grateful to A. Satou and D. Svintsov for discussions and useful comments. The work at RIEC  and UoA was supported by the Japan Society for Promotion of Science,
KAKENHI Grant   No. 16H06361. The work at RIEC with UB was supported by the RIEC Nation-Wide Cooperative Research Project. 
VR also acknowledges the support by  the Russian Scientific Foundation, Grant No.14-29-00277.
The work by VL was supported by the Russian Foundation for Basic Research, Grant No. 16-29-03402. The work at RPI were supported by  the US ARL Cooperative Research Agreement.

\end{document}